| RESEARCH ARTICLE

# Boosting Stock Price Prediction with Anticipated Macro Policy Changes

Md Sabbirul Haque[1] ✉ Md Shahedul Amin[2], Jonayet Miah[3], Duc Minh Cao[4] and Ashiqul Haque Ahmed[5]
[1]Institute of Electrical and Electronics Engineers, Piscataway, NJ 08854, USA
[2]Department of Finance & Economics, University of Tennessee, Chattanooga, TN, USA
[3]Department of Computer Science, University of South Dakota, South Dakota, USA
[4]Department of Economics, University of Tennessee, Knoxville, TN, USA
[5]Economics & Decision Science, University of South Dakota, South Dakota, USA
**Corresponding Author:** Md Sabbirul Haque, **E-mail**: sabbir465@ieee.org

| ABSTRACT

Prediction of stock prices plays a significant role in aiding the decision-making of investors. Considering its importance, a growing literature has emerged trying to forecast stock prices with improved accuracy. In this study, we introduce an innovative approach for forecasting stock prices with greater accuracy. We incorporate external economic environment-related information along with stock prices. In our novel approach, we improve the performance of stock price prediction by taking into account variations due to future expected macroeconomic policy changes as investors adjust their current behavior ahead of time based on expected future macroeconomic policy changes. Furthermore, we incorporate macroeconomic variables along with historical stock prices to make predictions. Results from this strongly support the inclusion of future economic policy changes along with current macroeconomic information. We confirm the supremacy of our method over the conventional approach using several tree-based machine-learning algorithms. Results are strongly conclusive across various machine learning models. Our preferred model outperforms the conventional approach with an RMSE value of 1.61 compared to an RMSE value of 1.75 from the conventional approach.

| KEYWORDS

Stock price forecasting, Machine Learning, Anticipated Macro Policy, macroeconomic variable

## 1. Introduction

Accurate forecasting of stock prices holds immense importance for investors and helps maintain a healthy portfolio, fostering enhanced profitability. An increasing number of investors and investment enterprises are now embracing advanced predictive models to enhance their portfolio management. In response, the academic realm has witnessed a surge in studies concentrating on predicting future stock prices. The current literature has harnessed various statistical methods along with advanced machine-learning approaches in an effort to achieve improved prediction accuracy.

Presently, the prevailing literature relies on time-series data related to stock prices, coupled with other relevant variables, to formulate forecasting models. However, it's crucial to acknowledge that overall economic conditions can significantly affect investments. Furthermore, some macroeconomic policy variables, such as interest rates, can be effectively anticipated ahead of time as the central bank frequently engages in public discussions regarding its future moves and strategies. Because interest rates can directly influence the cost of investment and thereby return from investments, investors have a strong incentive to form expectations about future interest rate changes ahead of time based on public discussions conducted by the central bank and accommodate their current investments based on their expected future interest rates. In our study, we incorporate future interest rates as a proxy for future expected interest rates along with other macroeconomic indicators, such as the Consumer Price Index (CPI), unemployment rates, and current interest rates, which can also impact the stock market significantly. To our knowledge, our study first tests this hypothesis and incorporates it into stock price prediction models for further performance improvement. Results from this study demonstrate that our novel approach outperforms conventional approaches. Specifically, we compare the performance of our method with that of conventional approaches using various machine learning algorithms. Results from our proposed method outperform conventional approaches in terms of the Root Mean Square Error (RMSE) for each machine learning algorithm explored in this study. Specifically, we report an RMSE value of 1.61 yielded by the best-performing model (LGBM) using our novel approach, whereas the conventional counterpart (LGBM without our proposed features) generates an RMSE value of 1.75. These findings have the potential to revolutionize investment opportunities by allowing superior predictions, thereby facilitating enhanced profitability.

## 2. Literature Review

The prior research in this field predominantly relies on classical approaches such as linear regression [Seber and Lee, 2013], linear time series models including Autoregressive Moving Average (ARMA) and Autoregressive Integrated Moving Average (ARIMA) [Zhang, 2003], Random Walk Theory (RWT) [Reichek and Devereux, 1982], Moving Average Convergence/Divergence (MACD) [Chong and Ng, 2008] to make predictions about stock prices. However, current research focuses on the use of advanced machine learning and artificial intelligence algorithms because of the enhanced performance of stock price prediction. Tree-based models such as Random Forest (RF) [Liaw & Wiener, 2002], as well as neural network-based approaches, such as Artificial Neural Networks (ANN), Convolutional Neural Networks (CNN), Recurrent Neural Networks (RNN), and Long Short-Term Memory (LSTM), have also been employed by researchers for stock prices prediction efforts [Li et al., 2017, Oyeyemi et al., 2007]. ANN possesses the ability to extract latent features through self-learning, making it an apt choice for stock price prediction. Its capacity as a strong approximator allows it to discern complex input-output relationships within extensive datasets. Thus, ANN emerges as a viable option for predicting an organization's stock prices.

Selvin et al. (2017) conducted stock price predictions for NSE-listed companies by comparing various deep-learning techniques. Hamzaçebi et al., 2009 explored multi-periodic stock market forecasting through methods like the ANN model. Rout et al. (2017) predicted the stock market utilizing a simplified RNN model, testing it on datasets from the Bombay Stock Exchange and the S&P 500 index. Roman et al. implemented RNN models on stock market data from five countries: Canada, Hong Kong, Japan, the UK, and the USA. These networks were trained to predict trends in stock returns [Roman and Akhtar, 1996]. Yunus et al., 2014 employed ANN on NASDAQ data to forecast stock closing prices. Mizuno et al. (1998) utilized ANN for technical analysis on the TOPIX dataset, applying it to a system for predicting buying and selling timing. Some studies have proposed the use of Random Forest (RF) for forecasting purposes. RF, an ensemble technique, excels in both regression and classification tasks. It constructs multiple decision trees during training, providing a mean regression of individual decision trees [Kumar and Thenmozhi, 1996]. Mei et al. (2014) effectively employed RF to predict real-time prices in the New York electricity market. Herrera et al. (2010) leveraged RF as a predictive model for forecasting hourly urban water demand. Khan et al. (2023) employ reinforcement learning algorithms to predict stock prices.

While there is a vast literature on stock price predictions, the literature fails to utilize the power of expected future macroeconomic policy changes as well as economic conditions and mostly focuses on historical stock price data for developing forecasting models. To our knowledge, no prior research has attempted to include macroeconomic conditions and expected macroeconomic policy changes in the predictive models for individual stock price predictions. Alamsyah and Zahir (2018) analyze macroeconomic variables such as inflation rates, interest rates, and exchange rates to forecast the IDX Composite Index, an index that measures the stock price performance of all listed companies on the Indonesia Stock Exchange. Haque (2023) takes into account such macroeconomic variables to forecast retail demands. Haque (2020) provides evidence for intertemporal behavioral adjustment for expected future fiscal policy changes. In this present study, we fill this gap and incorporate expected future macroeconomic policy changes as well as economic conditions along with historical stock prices to predict future stock prices. We first empirically demonstrate that future interest rates can influence stock price movements, and then we further demonstrate that prediction performances significantly improve by including our proposed features in the model.

## 3. Methodology

### *3.1 Data Preprocessing and Feature Engineering*

Data used in this study has been collected from multiple sources. Historical daily stock price data for S&P500 ticker symbols have been collected from publicly available Yahoo Finance. For these 500 ticker symbols, we calculate Partial Autocorrelations (PAC) for up to 59 lag values. We then find the maximum of 59 PAC values for each ticker symbol and rank ticker symbols based on that maximum PAC value. We choose the first ten ticker symbols in that ranked list. We select ticker symbols in this way to select those ticker symbols that have strong serial autocorrelation, which is essential for building a time-series model. Interest rates are collected from Fred's data repository. Furthermore, unemployment rates and Consumer Price Index (CPI) data have been collected from the World Bank's World Development Indicators (WDI) database. These individual datasets are then merged into a single data set. We append these datasets into a single dataset.

Table 1: List of Independent Variables

| Independent Variables | Descriptions |
| --- | --- |
| volume | Number of stocks traded |
| cpi | Consumer Price Index |
| unemp | Unemployment rates |
| int_rate | Current interest rates |
| lead_t7_int_rate | Expected future interest rates of 7 days ahead |
| lead_t14_int_rate | Expected future interest rates of 14 days ahead |
| lead_t21_int_rate | Expected future interest rates of 21 days ahead |
| lead_t28_int_rate | Expected future interest rates of 28 days ahead |
| lag_t28 | Lagged value of 28 days of stock price |
| rolling_mean_t7 | 7 days rolling average of 28 days lagged values of stock price |
| rolling_mean_t30 | 30 days rolling average of 28 days lagged values of stock price |
| rolling_mean_t60 | 60 days rolling average of 28 days lagged values of stock price |
| rolling_mean_t90 | 90 days rolling average of 28 days lagged values of stock price |
| rolling_mean_t180 | 180 days rolling average of 28 days lagged values of stock price |
| rolling_std_t7 | 7 days rolling standard deviation of 28 days lagged values of stock price |
| rolling_std_t30 | 30 days rolling standard deviation of 28 days lagged values of stock price |
| Indicator variables | Indicators for months of a year, days of a week, and ticker symbols |

Certain feature engineering has been conducted to facilitate desired results. We create a marker for each month of the year and day of the week. We include features for lead interest rates as a proxy for expected future interest rates. We include 7, 14, 21, and 28 leads for future interest rates. We also include 7, 30-, 60-, 90- and 180-day rolling averages of 28-day lagged values of stock price. I restrict the analysis to the years 2017-2019 to avoid capturing any trends that may not exist in recent years. All variables included in the model are presented in Table 1.

### 3.2. Machine Learning Algorithms

First, we empirically validate that current and future interest rates, along with other economic indicators, are associated with stock prices. To validate that, we estimate a linear regression of all the features on stock prices and report relevant co-efficient. After we verify the relevancy of our proposed variables in explaining stock prices, the next step is to train several machine learning algorithms with and without our proposed variables. Various machine learning models that we train include the Light Gradient Boosting Model (LGBM), Extreme Gradient Boosted Model (XGBM), and Decision Tree Model. We compare performances of each machine learning model trained on each dataset: with and without proposed variables.

### 3.2.1 Decision Tree Model

Decision Tree Regression model is a tree-based algorithm where the entire predictor space is split into a number of prediction regions. The predicted value for any observation within that region is then calculated by the mean or mode of all observations within that region. The predictor space is typically split into segments and can be represented as a tree, which consists of several splitting rules. The algorithm can be summarized below.

1. First, we split the predictor space—into J distinct and non-overlapping regions, $R_1$, $R_2$, . . . , $R_J$. We choose the predictor and cut point such that the resulting tree has the lowest RSS.

2. For every observation that falls into the region $R_j$, the response value is the same, the mean of all observed response values within that region.

### 3.2.2 Extreme Gradient Boosting Model

Extreme Gradient Boosting (XGB) is an implementation of a gradient-boosting decision tree algorithm that attempts to accurately predict a target variable by combining the estimates of a set of simpler and weaker models. It prevents overfitting and penalizes more complex models by introducing LASSO (L1) and Ridge (L2) regularization. The objective function is comprised of two parts: the first part represents the deviation of the model and is measured by the difference between the predicted value and the actual value, and the second part is the regularization term. The prediction accuracy of the model is determined by the deviation and variance of the model. The training process takes place iteratively, adding new trees that predict the residuals of prior trees that are then combined with previous trees to make the final prediction.

*Boosting Stock Price Prediction with Anticipated Macro Policy Changes*

### 3.2.3 Light Gradient Boosting Machine

The Light Gradient Boosting Machine (LGBM) algorithm is a variant of Gradient Boosting Decision Tree, which expands vertically, i.e., leaf-wise, while other algorithm trees expand horizontally. It approximates loss functions with second-order Taylor approximation at each step and then trains a decision tree to minimize the second-order approximation. LGBM improves efficiency over other Decision Tree models and relaxes the necessity of scanning all possible data points by introducing two novel techniques: Gradient-based One-Side Sampling (GOSS) and Exclusive Feature Bundling (EFB). GOSS allows us to exclude a significant proportion of data instances with small gradients and use the rest to estimate the information gain. On the other hand, EFB allows us to bundle mutually exclusive features to reduce the number of features. With LGBM, we can effectively reduce the number of features without hurting the accuracy of the split point.

### 4. Results and Discussions

In order to examine the relationship between stock prices and current as well as future interest rates along with other macroeconomic variables, we estimate a linear regression of all features on stock prices. Table 2 presents estimates from this regression. Coefficients on several other variables are suppressed for brevity. In this table, coefficients on macroeconomic variables, along with expected future interest rates, are presented along with their p-values. Coefficients on sale volume, CPI, unemployment rates, current interest rates, and expected future interest rates of 28 days ahead are statistically significant at a 10% level of significance. The coefficient on volume is -4.87E-07 and is statistically significant, implying a higher sale volume is associated with a lower stock price. The coefficient on CPI is 4.5751, which is significant at any level of significance, implying that a higher inflation rate is associated with higher stock price. Similarly, according to estimated results, higher unemployment rates have a statistically significant positive association with stock prices. The coefficient on current interest rates is negative and is statistically significant at any level of significance. This finding is consistent with the intuition that as the interest rate increases, investment becomes less profitable and less attractive, thereby causing the stock price to go down. Coefficients on all future expected interest rates are positive, with the coefficient on the 28-day expected future interest rate being statistically significant, whereas coefficients on other expected variables are not. These results are also consistent with the intuition that as investors anticipate that the cost of investment is going to rise in the future, it creates incentives for them to invest today before an actual interest rate hike takes place.

Table 2: Estimates for regression on stock prices

|  | coef | p-value |
|---|---|---|
| Volume | -4.82E-07 | 0.000 |
| cpi | 4.5751 | 0.000 |
| unemp | 10.5741 | 0.012 |
| int_rate | -17.7971 | 0.000 |
| lead_t7_int_rate | 0.4285 | 0.600 |
| lead_t14_int_rate | 0.8374 | 0.322 |
| lead_t21_int_rate | 0.0552 | 0.064 |
| lead_t28_int_rate | 3.4171 | 0.000 |

Co-efficient on Several other variables are suppressed for brevity.

Results from Table 2 provide empirical evidence that all proposed economic variables, along with expected macroeconomic policy variables, have a statistically significant relation with the stock price. These results justify the inclusion of these proposed variables into the predictive models. To empirically verify and compare the performance of predictive models, we train machine-learning models, such as LGBM, XGB, and Decision Tree, using two different datasets: one with these proposed variables and another without proposed variables. Detailed results are presented in Table 3.

Table 3: Comparative Performance of Models

|  | Model | Without proposed variables | | With Proposed variables | |
|---|---|---|---|---|---|
|  |  | RMSE | MAE | RMSE | MAE |
| 1 | Light GBM | 1.751 | 1.249 | 1.652 | 1.182 |
| 2 | XGB Regressor | 1.748 | 1.176 | 1.615 | 1.105 |
| 3 | Decision Tree | 1.765 | 1.024 | 1.738 | 1.070 |

In Table 3, we evaluate the performance of models using Root Mean Square Error (RMSE) and Mean Absolute Error (MAE). We train three tree-based machine-learning models using two datasets: with and without our proposed variables. Results from all three models trained on a dataset that includes our proposed variables display superior performance compared to that trained on a dataset without such variables in terms of both RMSE and MAE, with one exception. The Light GBM model fitted on data without our proposed variables generates an RMSE value of 1.751, whereas that trained on data with those variables displays an RMSE value of 1.652. MAE is also smaller when we incorporate our proposed variables. XGB also demonstrates superior performance when we include our proposed variables in terms of both evaluate metrics. For the decision tree model, our method provides superior performance in terms of RMSE value but inferior performance in terms of MAE measure. In general, these results strongly support our claim that stock price prediction performances can be significantly improved by including macroeconomic variables that can influence the stock market along with expected future macro policy variables. These results are consistent with our initial findings that our proposed variables are associated with stock prices and, therefore, demonstrate the promise of explanatory power in explaining stock prices.

## 5. Conclusion

In this research, we present an innovative approach to forecast stock prices employing machine learning algorithms. Specifically, we incorporate macroeconomic indicators and anticipated future shifts in macroeconomic policies into the machine learning models. We initially provide empirical evidence that these anticipated policy changes and other macroeconomic factors are pertinent in predicting stock prices. Subsequently, we compare the performance of each model trained on two distinct datasets: one with and one without the inclusion of our suggested variables. We assess model performance using two metrics: RMSE and MAE. Our results are robust across machine learning models (with the exception of the MAE for the Decision Tree model) and provide compelling evidence that our methodology outperforms existing approaches. These outcomes hold significant potential for both academic research and practical industry applications. This proposed technique has the potential to disrupt the investment market, as the results directly correlate with enhanced return on investment. Investors can employ our suggested method to make more informed and advantageous investment choices. In our research, we analyzed and contrasted the predictive model performances based on anticipated policy changes spanning one, two, three, and four weeks. Subsequent studies could delve into the sensitivity of this timeframe in relation to predictive accuracy.

**Funding:** This research received no external funding.
**Conflicts of Interest:** The authors declare no conflict of interest.
**Publisher's Note**: All claims expressed in this article are solely those of the authors and do not necessarily represent those of their affiliated organizations, or those of the publisher, the editors and the reviewers.